\documentclass[epsf,psfig]{mn2e}
\usepackage{amsmath, amsfonts, amssymb,epsfig}

\begin{document}
\renewcommand{\baselinestretch}{1}

\newcommand{\bnabla}{{\boldsymbol{\nabla}}}
\newcommand{\bu}{\mathbf{u}}
\newcommand{\bB}{\mathbf{B}}
\newcommand{\be}{\mathbf{e}}

\def\msun{{\rm M}_\odot}
\def\rsun{{\rm R}_\odot}
\def\halpha{H$\alpha$}
\def\hbeta{H$\beta$}
\newcommand{\lsimeq}{\mbox{$\, \stackrel{\scriptstyle <}{\scriptstyle
\sim}\,$}}
\newcommand{\gsimeq}{\mbox{$\, \stackrel{\scriptstyle >}{\scriptstyle
\sim}\,$}}
\def\gradi{\ifmmode{^\circ}\else$^\circ$\fi}
\def\reference{\parskip 0pt\par\noindent\hangindent 0.5 truecm}

\def\kms{km ${\rm s}^{-1}$}

\title[Fossil magnetic fields]{Modelling of isolated radio pulsars and magnetars 
on the fossil field hypothesis}

\author[L. Ferrario \&  D. T. Wickramasinghe]
{Lilia Ferrario and Dayal Wickramasinghe \\
Department of Mathematics, The Australian National University,
Canberra, ACT 0200, Australia}

\date{Accepted. Received ; in original form}

\def\Msun{{\rm M}_\odot}
\def\rsun{{\rm R}_\odot}
\def\lsun{{\rm L}_\odot}
\def\gradi{\ifmmode{^\circ}else$^\circ$\fi}
\def\reference{\parskip 0pt\par\noindent\hangindent 0.5 truecm}

\maketitle
\begin{abstract}

We explore the hypothesis that the magnetic fields of neutron stars are of
fossil origin. For parametrised models of the distribution of magnetic flux on
the Main Sequence and of the birth spin period of the neutron stars, we
calculate the expected properties of isolated radio pulsars in the Galaxy
using as our starting point the initial mass function and star formation rate
as a function of galacto-centric radius.  We then use the 1374 MHz Parkes
Multi-Beam Survey of isolated radio pulsars to constrain the parameters in our
model and to deduce the required distribution of magnetic fields on the main
sequence.  We find agreement with observations for a model with a star
formation rate that corresponds to a supernova rate of 2 per century in the
Galaxy from stars with masses in the range $8 - 45\Msun$ and predict 447,000
active pulsars in the Galaxy with luminosities greater than 0.19
mJy~kpc$^2$. The progenitor OB stars have a field distribution which peaks at
$\sim 46$~G with $\sim 8$ percent of stars having fields in excess of 1,000~G. The
higher field progenitors yield a population of 24 neutron stars with fields in
excess of $10^{14}$~G, periods ranging from 5 to 12 seconds, and ages of up to
100,000 years, which we identify as the dominant component of the magnetars.
We also predict that high field neutron stars ($\log B>13.5$) originate
preferentially from higher mass progenitors and have a mean mass of
$1.6\Msun$, which is significantly above the mean mass of $1.4\Msun$
calculated for the overall population of radio pulsars.

\end{abstract}


\section{Introduction}

A striking feature of the white dwarfs and the neutron stars is the wide range
of magnetic field strengths seen in each of these groups. In isolated magnetic
white dwarfs (about 15 percent of the total white dwarf population), the field
strengths are measured directly through the Zeeman effect and range from $\sim
10^5 - 10^9$~G.  In isolated radio pulsars, the field estimates are in the
range $\sim 10^{11} - 10^{14}$~G and are based on the measured spin down rates
and the assumption of dipole radiation. Observations in the X and $\gamma$-ray
spectral regions have revealed the presence of a class of neutron stars with
higher inferred fields, namely the Anomalous X-ray Pulsars (AXPs) and the Soft
Gamma Repeaters (SGRs). These neutron stars are often referred to as
``magnetars'' and have estimated field strengths in the range $\sim10^{14} -
10^{15}$~G.

Possible explanations for the magnetic fields in neutron stars are that the
fields are either (i) essentially of fossil origin (e.g. Rudermann 1972) or
(ii) generated by a convective dynamo (Thompson \& Duncan 1993). At the
present time, this issue remains unresolved (see Mestel, 1999, for a
comprehensive review).  In the case of the white dwarfs, a direct link can be
made between magnetism in the compact star phase and the Main Sequence via the
well documented properties of the chemically peculiar Ap and Bp stars
(Wickramasinghe \& Ferrario 2005) and has been a major factor supporting (i)
for the white dwarfs.

Magnetism is not as well documented in the more massive ($\gsimeq 8 \Msun$)
upper Main Sequence stars that evolve into neutron stars. The only known stars
with a directly measured magnetic field are $\theta$ Orion C, with a dipolar
magnetic field $B_p = 1,100$~G (Donati et al. 2002), and HD191612 with a field
of $B_p = 1,500$~G (Donati et al. 2005).  Their mass is estimated around $40
\Msun$, with the latter star slightly more evolved than the
former. Interestingly, the magnetic fluxes of both these stars ($1.1\times
10^{27}$ G~cm$^2$ for $\theta$ Orion C and $7.5\times 10^{27}$ G~cm$^2$ for
HD191612) are comparable to the flux of the highest field magnetar SGR~1806-20
($5.7\times 10^{27}$ G~cm$^2$). There is also indirect evidence from studies
of stellar winds from massive stars that magnetism may be widespread among OB
stars (e.g. Wade 2001). These facts, taken together with the scaling that
exists between the dipolar field strengths of the white dwarfs and the neutron
stars (Figure 1), suggest that a prima facie case can be made for a fossil
origin of fields also in the neutron stars. The fossil field hypothesis
provides a natural explanation for the wide range of observed magnetic field
strengths in a given class as being the result of inhomogeneities in the
interstellar magnetic field strength in star forming regions.

Calculations of pre-Main Sequence evolution have shown that stars more massive
than $\sim 2\msun$ are likely to begin their Main Sequence phase with a
primordial magnetic flux entrapped in radiative regions supporting the fossil
hypothesis (Tout, Livio \& Bonnell 1999; Moss 2003).  There are no detailed
studies on the interplay between fossil fields and subsequent stellar
evolution. This problem has been discussed by Tout, Wickramasinghe \& Ferrario
(2004, hereafter TWF). They argued that provided there are regions of the star
that are radiative during subsequent evolution, a fossil Main Sequence poloidal
magnetic flux could survive through to the compact star phase.

\begin{figure}
\begin{center}
\hspace{0.1in}
\epsfxsize=0.98\columnwidth
\epsfbox[30 154 577 512]{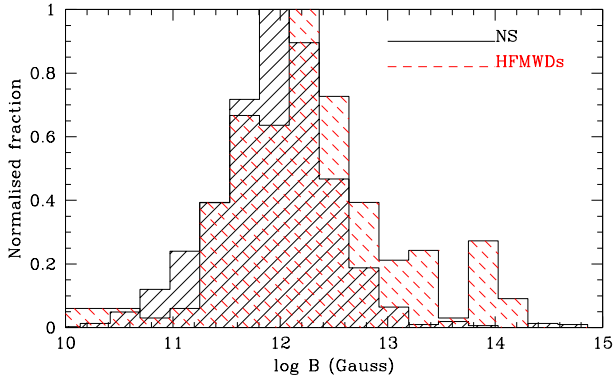}
\caption{Black histogram (solid line): Field distribution of neutron
stars. Red histogram (dashed line): Field distribution of High Field Magnetic
White Dwarfs when their radii are shrunk to that of a typical neutron star
radius of $\sim 10^6$~cm.}
\end{center}
\end{figure}

In this paper, we investigate the consequences of the fossil field hypothesis
for the origin of magnetic fields in neutron stars by carrying out population
synthesis calculations for different assumptions on the distribution of the
magnetic flux of massive ($\gsimeq 8 \Msun$) Main Sequence stars and of the
field dependence of the initial birth period of neutron stars. We model the
observed properties of the population of isolated radio pulsars in the 1374
MHz Parkes Multi-Beam Survey (PMBS, Hobbs et al. 2004; Kramer et al. 2003;
Morris et al. 2002; Manchester et al. 2001) to constrain our model parameters,
and use these to deduce the required magnetic properties of the progenitor OB
stars.

\section{The model}

\subsection{The progenitors}

We assume that neutron stars are formed in the galactic disc from stars with
$8\leq M/\msun \leq 45$ and that black holes form from larger mass stars. The
situation is not as clear cut, from a theoretical point of view, because of
the uncertainties in estimating the effects of fall-back (e.g. Heger et
al. 2003). Nonetheless, there is now overwhelming observational evidence that
progenitors of neutron stars could be as massive as $40-50\msun$
(e.g. Gaensler et al. 2005; Muno et al. 2005).

We assume an initial mass function $\psi(M_i)$ with a power law index
$\alpha_3=2.7$ (Kroupa 2002), and a constant star formation rate $S(r_g)$ at a
given galacto-centric radius $r_g$. The radial dependence of $S(r_g)$ is
obtained from the data presented by Boissier et al. (2003).  The stars are
distributed perpendicularly to the galactic plane (the $z$ direction)
exponentially using the scale heights as a function of initial mass given by
Rana (1987) supplemented at the high mass end by the observed Wolf-Rayet
distribution of Conti \& Vacca (1990). These values range from a scale height
of 100 pc for a $8.4\Msun$ star to 45 pc for a $45 \Msun$ star, which is in
agreement with the more recent results of Reed (2000).

Unlike in the case of the white dwarfs, for the neutron stars the initial
$(M_i)$ to final mass $(M_{\rm NS})$ relationship is not well constrained by
observations. Masses are known for a handful of neutron stars, particularly in
binaries, and these are observed to cluster around $M_{\rm NS} \sim 1.4 \msun$,
close to the core collapse mass for an initial mass $M_i \sim 8 \msun$. In our
calculations, we use the relationship obtained by Heger, Woosley \& Spruit (2005).

In the simple model of TWF, a star of mass $M_i$ that collapses from the
interstellar medium entraps a magnetic flux that is proportional to the
ambient interstellar (primordial) magnetic field $B_{\rm ISM}$ and the square
of its initial radius $R_i$. If we assume an initially nearly uniform density,
it follows that the magnetic flux $\Phi \propto B_{\rm ISM} M_i^{2/3}$. The
variations expected in the interstellar magnetic field will then determine the
distribution of magnetic fluxes in the progenitor stars.  We parametrise the
magnetic flux distribution $\chi(\Phi)$ by assuming that all massive stars are
magnetic, and that the magnetic flux $\Phi$ is given as the sum of two
Gaussians in the logarithm with dispersions $\sigma_{\log\Phi1}$ and
$\sigma_{\log\Phi2}$ with appropriate weightings. The mean values are assumed
to have the form
\begin{equation}
\label{flux} 
\langle\log \Phi\rangle=\log\left(\Phi_{m}\right)+ \frac{2}{3}\log\left(\frac 
{M_i}{9 \Msun}\right)
\end{equation}
where the subscript $m$ takes the values 1 and 2 for the two Gaussians.
The magnetic fluxes are in units of G~cm$^2$.
\subsection{Parametrisation of the birth properties of the neutron stars}

With the prescription of \S 2.1, we generate neutron stars with a range of
masses. Our assumption that the magnetic flux is conserved during post main
sequence evolution to the compact star phase allows us to calculate the birth
surface magnetic fields of these neutron stars from:
\begin{equation}
\label{Bf} 
B_{\rm NS}=\frac{\Phi}{\pi (R_{\rm NS})^2}~~ {\rm G}
\end{equation}
where $R_{\rm NS}$ is the radius of the neutron star in cm given by the
mass-radius relationship for neutron stars which depends on the equation of
state.  We adopt the values obtained with the unified SLy equation of state of
Douchin \& Haensel (2001).

We follow the motions of these stars by integrating the equations of motion in
the galactic potential of Kuijken \& Gilmore (1989) assuming that they are
born with a kick velocity that is independent of the progenitor mass and is
given by a Gaussian distribution (e.g. Hobbs et al. 2005) with velocity
dispersion $\sigma_v$.

Radio pulsars are generally observed not to have a braking index of 3 (e.g.
Cordes \& Chernoff 1998).  However, to minimise the number of free parameters
we assume, in common with many other theoretical studies, that the spin down
of radio pulsars is by dipole radiation:
\begin{equation}
\label{pdotnow}
\dot P= 9.76\times 10^{-16}\!\!\left(\frac{R_{\rm
NS}}{10^6\rm{cm}}\right)^6\!\!\left(\frac{10^{45}\rm{g}\,\rm{cm}^2}{I}\right)\!\!
\left(\frac{B_{\rm NS}}{10^{12}\rm{G}}\right)^2\!\!\left(\frac{\rm s}{P}\right)~{\rm s s}^{-1}
\end{equation}
where $I$ is the moment of inertia as derived by Douchin \& Haensel (2001) and
$P$ is the period in seconds. We have explicitly assumed that the time scale 
for field decay is longer than the ages of the pulsars, although 
the issue of whether field decay is necessary to explain their detailed  
properties remains an open question (e.g. Gonthier et al. 2002). 

We may expect the birth period of the compact star to be determined by the
magnetic flux of the progenitor star and its initial angular momentum
(e.g. Brecher \& Chanmugam 1983; Narayan 1987). The complex interactions that
occur between magnetic fields and rotation during stellar evolution and the
mechanisms responsible for the transport of angular momentum are still poorly
understood (e.g. Spruit \& Phinney 1998; Watts \& Andersson 2002; Fryer \&
Warren 2004; Heger et al. 2005; Ott et al. 2005). Magnetic fields must play a
role in transporting angular momentum outwards from the stellar core to the
stellar envelope because otherwise all compact stars would be born at break up
velocity, which is not the case (e.g. Migliazzo et al 2002 and Kramer et
al. 2003).  Heger et al. (2005) have considered this problem, but only in the
weak field regime not directly relevant to the fossil field hypothesis. They
report that the rotation periods they find are generally too small to explain
pulsar observations. On the other hand, Ott et al. (2005) have proposed that a
propeller mechanism operating on fall-back material may slow down the rotation
of pulsars at birth. Interestingly, we know that there is strong direct
evidence for higher field objects to be slower rotators in the magnetic white
dwarfs and we speculate that this may be a general characteristic of stars
with magnetic fields of fossil origin (see Ferrario \& Wickramasinghe 2005 for
scaling of rotation properties) and thus allow for this possibility in our
modelling.

We model the birth spin periods of neutron stars $P_0(B_{\rm NS})$ by assuming
they are distributed as a Gaussian $\Pi(P_0)$ with a dispersion $\sigma_{P_0}$
about a mean $P_{0}(B_{\rm NS})$ which depends linearly on the magnetic field

\begin{equation}
\label{P0m}
\langle P_{0}(B_{\rm NS})\rangle=\left(\frac 
{P_a}{s}\right)+\alpha\left(\frac{B_{\rm
NS}}{10^{12}\rm{G}}\right)~~ {\rm s}
\end{equation}
This is clearly a first approximation (e.g. in fall back models, a field
dependence $\propto B_{\rm NS}^{6/7}$ may be more appropriate). For different
assumed values of $\alpha\ge 0$, we constrain the parameters $P_a$, and
$\sigma_{P_0}$ by fitting the observations of isolated radio-pulsars in the
PMBS taken from the ATNF catalogue (Manchester et al. 2005).

To fit these observations, we also have to model the radio luminosity.  Guided
by the expected power law dependence on $P$ and $\dot P$, and following
previous investigators, we have assumed that the luminosity $L_{400}$ at
400~MHZ can be described by a mean luminosity of the form
\begin{equation}
\label{lm}
\log \langle L_{400}\rangle=\frac{1}{3}\log\left(\frac{\dot
  P}{P^3}\right)+\log L_0
\end{equation}
with $\log L_0=7.2$.  Here the luminosities are in units of mJy kpc$^2$. We
have modelled the spread around $L_{400}$ using the dithering function of
Narayan \& Ostriker (1990).
\begin{equation} 
\label{gamfun}
\rho_L(\lambda)=0.5\lambda^2\exp\left(-\lambda\right)\qquad\qquad (\lambda\ge 0)  
\end{equation}
where
\begin{equation}
\label{dith}
\lambda=b\left(\log\frac{L_{400}}{\langle L_{400}\rangle}+a\right)
\end{equation}
and $a$ and $b$ are constants to be determined (Hartman et al. 1997). The
luminosity at 400 MHZ is then scaled to the observed PMBS frequency of 1374
MHZ using a spectral index of $-1.7$ (Gonthier et al. 2002). 

In our studies, we have adopted the radio death line predicted for a
multipolar field configuration near the stellar surface in a
space-charge-limited flow model (Zhang, Harding \& Muslimov 2000), that is:
\begin{equation} 
\label{death}
\log \left(\frac{\dot P}{{\rm s s}^{-1}}\right)=2\log\left(\frac{P}{\rm 
s}\right)-16.52
\end{equation}

\subsection{The calculation of population characteristics}

We calculate the total number of neutron stars in the Galaxy with period up to
$P$ by evaluating the integral
\begin{eqnarray}
\displaystyle{N(\le P)}\!\!\!\!\!\! &=& 
\!\!\!\!\!\!\displaystyle{\int_{0}^{2\pi}\!\!\!\!
\int_{0}^{15}\!\!\!\!\int_{P_{01}}^{P_{02}}\!\!\!\!\int_{P_0}^P\!\!\int_{\Phi_{m
in}}^{\Phi_{max}}}\!\!\!\!\!\int_{8}^{45}\!\!\!\psi
(M_i)\,\chi(M_i,\Phi)\, \Pi(P_0)\,r_g S(r_g) \nonumber \\
 & & ~~~~~~~\times ~~  dM_i\, d\Phi\,\frac{dP}{\dot P}\,dP_0\,dr_g\,d\phi
\label{numb}
\end{eqnarray}
Here, $r_g$ (in kpc) is the galacto-centric radial coordinate, and $\phi$ is
the galactic longitude. $\dot P$ is calculated from equation \ref{pdotnow} in
appropriate units using the magnetic field $B_{\rm NS}(\Phi(M_i))$ derived from
equation \ref{Bf}. The limits $P_{01}$, $P_{02}$, $\Phi_{min}$, $\Phi_{max}$
are chosen to adequately sample the Gaussian distributions.

To compare our theoretical calculations with the PMBS isolated pulsars, we
need to know the current location of each star that can contribute to the
integral. To achieve this, we distribute the stars in velocity space according
to the prescription of \S 2.2 with $\sigma_v=380$ km s$^{-1}$ and in the $z$
direction according to their progenitors masses as detailed in \S 2.1. Their
spatial location at the current epoch is then determined by solving the
equations of motion in the galactic potential. The dispersion and scattering
measures {\it DM} and {\it SM} of the pulsars are calculated using the FORTRAN program
of Cordes \& Lazio (2002). Finally, the pulsar luminosities are assigned
according to equations \ref{lm} to \ref{dith}.

Once all the intrinsic properties of our model pulsars are determined, these
are fed through a filtering program kindly provided to us by Natasa Vranesevic
(2005, private communication) that checks the pulsars for detectability by the
PMBS. 

Finally, since pulsars radio emission is anisotropic, we use the Tauris
\& Manchester (1998) beaming model to obtain the correct fraction of detected
pulsars:
\begin{equation}
\label{beam}
f(P)=0.09\left[\log\left(\frac{P}{{\rm s}}\right)-1\right]^2+0.03
\end{equation}

\section{Results and discussion}

Our approach has been to focus on constraining the key parameters that we have
introduced for describing the magnetic flux distribution on the Main Sequence
and the distribution of the initial birth period assuming that the remaining
parameters of the model are within the ranges given by previous investigators.
We have used as our basic observational constraints the 1-D projections of the
data comprising the number distributions in period $P$, magnetic field
$B_{NS}$, period derivative $\dot P$ and radio luminosity $L_{1374}$, noting
that these distributions are not all independent. 

\begin{figure*}
\begin{center}
\hspace{0.1in}
\epsfxsize=0.8\textwidth
\epsfbox[18 439 570 700]{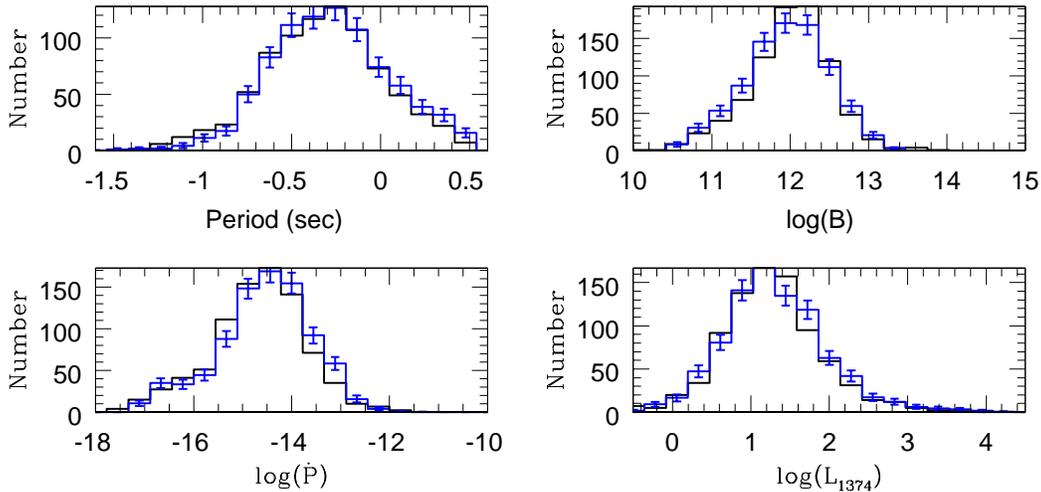}
\caption{Our model (blue) overlapped to the PMBS isolated radio pulsars data
from the ATNF catalogue (http://www.atnf.csiro.au/research/pulsar/psrcat). The
errors are given by $\sqrt{N}$, where $N$ is the number of objects in each
bin.}
\end{center}
\end{figure*}

The best fit model to the observations of the PMBS pulsars was determined ``by
eye'' after conducting hundreds of trials. We have found this method more
reliable than the standard K-S statistic. Our results are shown in Figure 2.

The initial period distribution is described by $P_a= 0.22$~s,
$\sigma_{P_0}=P_0/2$ and $\alpha = 0.01$.  The magnetic flux distribution is
described by two Gaussians of means and spreads $\log\Phi_1=26.2$,
$\sigma_{\log\Phi1}=1.0$, and $\log\Phi_2= 25.2$, $\sigma_{\log\Phi2} =0.61$
respectively, weighted in the ratio $1:5$. The parameters for the luminosity
model are $a=1.8$ and $b=2.8$.

This model reproduces the observed total number of pulsars in the PMBS, once
we exclude the millisecond pulsars ($P<20$~ms). The star formation rate is
normalised to $2.5\Msun$ pc$^{-2}$ Gyr$^{-1}$ at the galacto-centric radius of
the Sun, which is within the uncertainty in the Boissier et al. (2003)
relationship. This corresponds to a galactic supernova rate of 2 per
century. 

In a broad sense, the distribution that we use for the magnetic flux has the
strongest influence on the location of the peak in the magnetic field
distribution of the pulsars, while the mean period $P_a$ and the width of the
period distribution $\sigma_{P_0}$ have a strong influence on the period
distribution, although clearly period and field distributions are inter-linked
through period evolution which depends on the magnetic field. We find that we
cannot model both the field distribution and the period distribution at the
same time if we assume that all pulsars are born at a period less than
$50$~ms. Such models give an excess of stars at low periods which is not
consistent with the results of Vranesevic et al. (2004) who find that up to 40
percent of pulsars are born with periods in the range 0.1-0.5~s. Regimbau \&
Freitas Pacheco (2001) also noted that most pulsars are born with $P>0.1$~s
and conduct their population synthesis studies using a Gaussian with a mean
birth period $P_0=0.29$~s and a spread of 0.1~s while Faucher-Gigu\`ere \&
Kaspi (2005) use a Gaussian with a mean birth period $P_0=0.3$~s and a spread
of 0.15~s. All these results contradict the view that all pulsars are born as
fast rotators ($P_0\le 0.1$~s) which arises from observational selection effects,
since young and fast neutron stars (e.g. the Crab pulsar) would be more easily
detectable as radio pulsars than the more slowly rotating ones.  We note that
with our model parameters, the pulsar birth spin period increases with
increasing magnetic field, but is almost independent of the field in the range
$\log B_{\rm NS}(G)=10-13$ where the vast majority of isolated radio pulsars
lie. The magnetic field dependence becomes important in the high-field radio
pulsars and magnetar field range.

The observations show that the field distribution is intrinsically asymmetric
with a low field tail that is more prominent than the high field tail. This
general behaviour is reproduced by our assumed bi-Gaussian flux
distribution. Furthermore, the model predicts 26 pulsars with fields $>
10^{13}$ G, in good agreement with the PMBS which shows 23 stars in the same
field range.

There have been many previous attempts at synthesising the population of radio
pulsars with different assumptions on neutron star magnetic fields, field
decay and initial periods.  Some of these investigations have led to the
conclusion that the neutron star field distribution cannot be described by a
single Gaussian (e.g. Gonthier et al. 2002), and we find a similar result for
the magnetic fluxes of their progenitor stars. 

The initial-final mass radius relationship that we have assumed implies that
higher mass progenitors will produce higher mass neutron stars. These stars
will, on average, also have higher magnetic fluxes, and therefore tend to
produce higher field neutron stars. However, the latter effect is
counterbalanced by the strongly declining initial mass function which favours
low mass star progenitors, which can also produce high field neutron stars in
the Gaussian tail of the magnetic flux distribution. In order to investigate
the mass distribution of active radio pulsars in the Galaxy with periods less
than 20~s, we have divided them into two field groups ($\log B_{NS}(G) = 10 -
13.5$) and ($\log B_{NS}(G) > 13.5$). Our results are shown in Figure 3 for
pulsars with 1374 MHZ luminosities $\ge 0.19$ mJy~kpc$^2$. The mean neutron star
masses in these groups are $1.45\Msun$ and $1.6\Msun$ respectively, whilst the
mean masses of their progenitors are $12 \msun$ and $19 \msun$
respectively. The stars in the highest field bin have a high mass tail with
54 percent of the stars extending beyond $1.5\Msun$ compared to 19 percent of stars in the
same mass range in the lower field bin. We find that the total number of
active pulsars with 1374 MHZ luminosities $>0.19$ mJy~kpc$^2$ is 447,000. If
we apply the beaming model of equation \ref{beam}, we find that the number of
active pulsars whose beams intercept Earth reduces to 48,000.

\begin{figure}
\begin{center}
\hspace{0.1in}
\epsfxsize=0.95\columnwidth
\epsfbox[36 313 379 642]{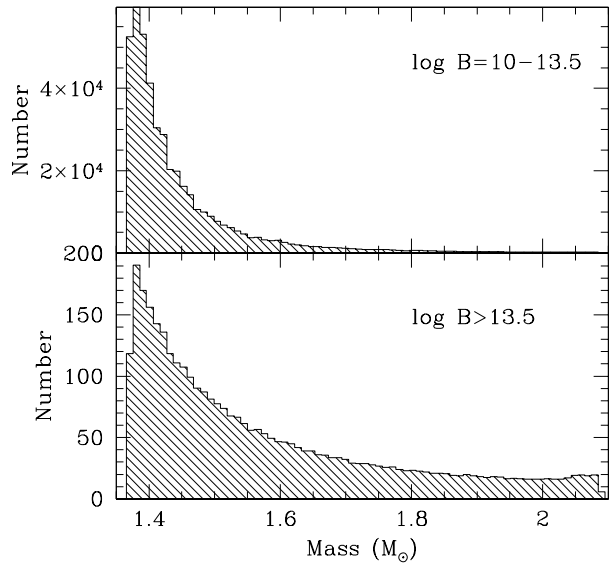}
\caption{Masses of neutron stars with rotational periods of up to 20 s in the
two indicated field ranges.}
\end{center}
\end{figure}
 
The magnetars are neutron stars that are discovered as X-ray and/or
$\gamma$-ray sources. They are currently confined to the period range $5 -
12$~s and are young objects (e.g. Gaensler et al. 2001). If one makes the
standard assumption of dipole radiation with a braking index of 3, the
magnetic fields are in the range $\log B_{\rm NS}(G) =13.8-15.3$.

We have used our calculations to investigate if the high field tail of the
neutron stars in our model can provide an explanation for the numbers of
magnetars that are seen in the Galaxy. Here we use as our cutoff $B_{\rm
NS}=10^{14}$~G, since this value is just above the revised quantum critical
field of Zhang \& Harding (2000) for radio emission. If we isolate the total
number of neutron stars with fields in excess of $10^{14}$~G, and ages of less
than 500,000 years, we find 146 such stars, none of which would be detectable
as radio-pulsars. If we now limit the age to 100,000 years, our modelling
yields 24 magnetars in the Galaxy with periods from $5$ to $12$~s and 8
magnetars with periods $>12$~s, 5 of which have periods between 12 and 15
seconds. Thus, the number declines rapidly with period outside the observed
period range. We note in this context that some magnetars exhibit a large
discrepancy between their real ages (as determined through their association
with supernova remnants) and their characteristic ages (e.g. AXPs 1E2259+586
and 1E1845-0545, Gaensler et al. 2001). This discrepancy is readily explained
through our assumption of a $P-B$ relationship, which sees high field neutron
stars being born preferentially at long periods ($1-10$~s).

Below $B_{\rm NS}=10^{14}$~G, there is an area of overlap where a couple of
magnetars have fields and periods that are comparable to those of the high
field radio pulsars (HBRPs) (e.g. Kaspi \& McLaughlin 2005).  This suggests
that there could exist lower field magnetars ($<10^{14}$~G) that eventually
evolve into X-ray silent HBRPs of the type currently observed if their
radio-beams intercept the Earth. We note that it should also be possible to
discover ``hybrid'' young objects with $B_{\rm NS}<10^{14}$~G exhibiting both
magnetar and radio pulsar characteristics. In our analysis, we have counted
all these objects as radio-pulsars if they pass through our PMBS filtering
program.

Finally, we return to the recent evidence for high progenitor masses in the
magnetars 1E~1048-5937 (Gaensler et al. 2005) and CXO~J164710.2-455216 (Muno
et al. 2005). In the dynamo model, rapid spin ($\sim 3$~ms) is required to
generate super-strong fields and Heger et al. (2005) argue that it is only the
cores of the more massive stars that have this property. This has been used as
an argument in favour of the dynamo model, at least for the magnetars.  We
note that these findings also support our fossil hypothesis, since we predict
that magnetars, and more generally high field neutron stars, originate
preferentially from high mass progenitors.

We conclude by noting that the fossil field hypothesis as formulated by us,
and which does not allow for magnetic flux loss in the post main-sequence
evolution, requires a very specific distribution of magnetic fields for
massive Main Sequence stars. This is shown in Figure 4 for an assumed dipolar
field structure. This distribution has a peak at $\sim 46$~G and low and high
field wings extending from $\sim 1$~G to $\sim 10,000$~G with 8 percent of stars
having fields in excess of $\sim 1,000$~G. The latter group are the progenitors
of the highest field magnetars. The detailed shape of the Main Sequence
distribution shown in Figure 4 is potentially an observable quantity, and a
prediction of our fossil field model.

\begin{figure}
\begin{center}
\hspace{0.1in}
\epsfxsize=0.98\columnwidth
\epsfbox[14 312 385 507]{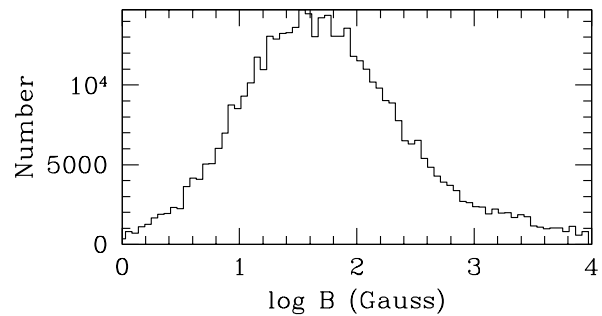}
\caption{Predicted magnetic field distribution of massive stars ($8-45\Msun$)
on the Main Sequence.}
\end{center}
\end{figure}

\section*{Acknowledgements}

We thank Natasa Vranesevic and Dick Manchester for providing us with the codes
necessary to compare our theoretical results with the observable sample of the
PMBS pulsars. We also thank the anonymous Referee for a careful reading of our
manuscript and for numerous useful comments.

\end{document}